\documentstyle[aps,prl,twocolumn,epsf]{revtex}

\begin{document}
\draft

\title{Theory of the field-induced gap in $S=1/2$ antiferromagnetic chains}

\author{Masaki Oshikawa$^1$ and Ian Affleck$^{1,2}$}

\date{22 May 1997}

\address{Department of Physics and Astronomy$^1$
and Canadian Institute for Advanced
Research$^2$, 
\\ University of British Columbia, Vancouver, BC, V6T 1Z1, CANADA}

\maketitle

\begin{abstract}
In a recent neutron-scattering experiment on
the quasi-one-dimensional $S=1/2$ antiferromagnet Cu Benzoate,
a gap was induced by an applied magnetic field.
We argue that the primary mechanism of the gap formation is
an effective staggered field due to both the alternating $g$-tensor
and the Dzyaloshinskii-Moriya interaction.
We explain the dependence of the gap on the applied field,
as well as identify several peaks in the structure factor $S(q,\omega)$. 
\end{abstract}

\pacs{PACS number: 75.10Jm}

Quantum spin chains have attracted much interest for a long time.
This is partly because sophisticated theoretical analysis, such as
exact solutions, can be applied to one dimensional systems.
Not only are they easier to analyze, it has been also recognized
that the effect of quantum fluctuations is more significant
than in higher-dimensional systems, resulting
in many interesting phenomena.
On the other hand, progress in experimental techniques has increased
the opportunity to observe physics of one-dimensional systems.
In a recent high field neutron-scattering
experiment~\cite{Dender:s12} on Cu Benzoate,
which is a (quasi-)one-dimensional $S=1/2$ antiferromagnet,
the field-induced shift in the soft-mode momentum is
observed for the first time.
Although the shift of the momentum is consistent with previous theoretical
analysis on the Heisenberg antiferromagnetic chain, the experiment also found
an unexpected excitation gap induced by the applied field.
The observed gap is proportional to ${H_0}^{0.65}$ where $H_0$ is the magnitude
of the applied field.
While the data is consistent with the power law with the same exponent
$0.65$ for three different directions of the applied field,
the coefficient depends on the direction.
The ratio of the coefficient is found to be $0.55:1.0:2.0$ for
the field applied in $a'',b,c''$ axes, which are the principal
axes of the effective exchange interaction.
(For detailed description of the compound,
see Ref.~\cite{Oshima:CuBenzAFR}.)
The observed gap can be as large as $0.3J$ where $J$ is the exchange
coupling in the chain direction, at $H_0 =7 {\rm T}$ where the average
magnetization is $0.06$ per site.

In this letter, we discuss the mechanism of the field-induced gap
observed in Cu Benzoate. We argue that the primary mechanism
is due to an effective staggered field.
As pointed out by Dender et al.~\cite{Dender:s12},
an effective staggered field is generated by the alternating $g$-tensor.
We found that the effective staggered field is also generated by
the Dzyaloshinskii-Moriya (DM) interaction, and the latter is no less
important than the former.
Our theory successfully explains the experimental data, including
the angle dependence of the gap.
For quasi-one-dimensional compounds with alternating crystal axes,
both effects are expected. Thus, our theory should apply to
such compounds in general.

Since the interchain coupling is very weak in the compound,
the gap formation should be understood primarily in a one-dimensional
model.
In this letter, we restrict our discussion to one-dimensional models.
As a first approximation, the system would be described as the
standard isotropic Heisenberg antiferromagnet.
Actually, the neutron scattering data at zero magnetic field
is consistent with the theoretical analysis based on the
standard Heisenberg model with exchange coupling
$J = 1.57 {\rm meV}$.
However, the standard Heisenberg model in an applied field
remains gapless from zero magnetic field up to saturated
magnetization~\cite{Griffiths}. Thus we have to consider some modification.

Even if we generalize the model Hamiltonian,
the system remains gapless for generic values of the magnetization,
as long as it has rotation symmetry about the direction of
the magnetization (we call this axial symmetry hereafter).
This can be seen from abelian bosonization or the generalized
Lieb-Schultz-Mattis theorem~\cite{s32lett}.
Since the gap is observed with a continuously changing magnetization,
it must be related to a breaking of the axial symmetry.
Let us discuss this using abelian bosonization.
Here we follow the notation and convention of Ref.~\cite{Affleck:LesHouches},
and take the direction of the magnetization as the quantization axis
($z$-axis.)
The breaking of the axial symmetry allows
a series of operators $e^{2 n \pi i R \tilde{\phi}}$, where $\tilde{\phi}$
is the dual field, $R$ is the compactification radius, and $n$ is an
integer.
If one of these operators appears in the Hamiltonian with a non-vanishing
coefficient, and it is relevant (in the Renormalization-Group sense),
we expect an energy gap.

A simple possibility of axial symmetry breaking is
the exchange anisotropy, including the dipole-dipole interaction.
However, the magnitude of the anisotropy is of order of
$1 \%$ of $J$~\cite{Oshima:CuBenzAFR,Dender:s12}.
From a bosonization analysis, which we do not discuss in detail in
this letter, we estimate the gap induced by the exchange anisotropy as
$10^{-5} J$ at the field of $7$ T.
This is too small compared to the experimental value up to $0.3J$.
Thus we must seek another mechanism.

Cu Benzoate has alternating crystal axes, which gives
an alternating $g$-tensor.
Due to the alternating $g$-tensor, a uniform applied field
produces an effective staggered field on the spin chain
as pointed out in Ref.~\cite{Dender:s12}.
Moreover, an additional contribution
to the effective
staggered field comes from the DM interaction,
which is also present due to the alternating crystal axes.
Both lead to a transverse staggered field, namely
the direction of the staggered field is (almost) orthogonal to the
direction of the magnetization.
As we will see, these two contributions to the staggered field
are of same order and both are important in analyzing the
angle-dependence of the gap.

According to Ref.~\cite{Oshima:CuBenzAFR},
the local $g$-tensor for Cu ions
is given by $g = {\rm diag}(2.08,2.05,2.36)$
in the local principal coordinates.
Due to the alternating direction of the oxygen octahedra around
the Cu ions,
the principal axes of the $g$-tensor alternates along the chain.
In the experiment, the field is applied in the principal directions
($a'',b,c'')$ of the total exchange anisotropy.
(For details, see Ref.~\cite{Oshima:CuBenzAFR}.)

The $g$-tensor in $a'', b, c''$-bases~\cite{Oshima:CuBenzAFR} is given by
\begin{equation}
 g = \left( \begin{array}{ccc}
              2.115 & \pm 0.0190 &  0.0906 \\
             \pm 0.0190 & 2.059 & \pm 0.0495 \\
             0.0906 & \pm 0.0495 & 2.316
            \end{array} \right),
\end{equation}
where $\pm$ corresponds to the two inequivalent sites.
For example, if we apply the magnetic field in $c''$-direction,
the effective staggered field generated by $g$-tensor is $(0, 0.025,0) H$
in $a''bc''$-coordinates. (The sign of the staggered field is defined
by referring to the even sites.) 
For field applied in $b''$ and $a''$ directions, it is
$(0.0095,0,0.025)H$ and $(0,0.0095,0)H$, respectively.

On the other hand, ignoring other than the nearest-neighbor interaction,
DM interaction in the chain can be written as
\begin{equation} 
  H_{\rm DM} = \sum_j (-1)^j \vec{D} \cdot ( \vec{S}_j \times \vec{S}_{j+1} ).
\end{equation}
Note that the factor $(-1)^j$ is present, as required from the crystal
structure~\cite{Moriya:DMint,Oshima:CuBenzAFR}.
When a magnetic field is applied,
an effective staggered field is generated through the DM 
interaction. 
While it is possible to see this by a Mean-Field argument,
it can be deduced from the following exact mapping.
In fact, we can eliminate the DM interaction
by a redefinition of the spin variables~\cite{RussianDM,Shekhtman:DMint}.
For simplicity, let us assume the $\vec{D}$ points in the $z$-direction.
Then, the Heisenberg Hamiltonian with a DM interaction
is given by
\begin{eqnarray}
  H &=& \frac{1}{2} \sum_j [ {\cal J} S^+_{2j-1} S^-_{2j}
                          + {\cal J}^*  S^+_{2j} S^-_{2j+1}
                          + (\mbox{h.c.}) ]
\nonumber \\
&&     + J \sum_j [ S^z_{2j-1} S^z_{2j} + S^z_{2j} S^z_{2j+1}] ,
\end{eqnarray}
where $S^{\pm} = S^x \pm i S^y$ and ${\cal J} = J + i D$.
By the rotation about $z$-axis by an alternating angle
\begin{equation}
  S^+_{2j} \rightarrow S^+_{2j} e^{ i \alpha /2} ,
  S^+_{2j-1}  \rightarrow S^+_{2j-1} e^{ - i \alpha/2},
\label{eq:redef}
\end{equation}
where $\tan{\alpha} = D/J$, the Hamiltonian is transformed to
\begin{eqnarray}
  H &=& \frac{1}{2} |{\cal J}| \sum_j [  S^+_{2j-1} S^-_{2j}
                          +  S^+_{2j} S^-_{2j+1}
                          + (\mbox{h.c.}) ]
\nonumber \\
&&     + J \sum_j [ S^z_{2j-1} S^z_{2j} + S^z_{2j} S^z_{2j+1}].
\end{eqnarray}
Namely, the DM interaction is eliminated,
resulting in an anisotropic exchange coupling.
This anisotropy would be cancelled by the exchange anisotropy
before the redefinition of eq.~(\ref{eq:redef}),
under some assumptions~\cite{Shekhtman:DMint}.
In any case, the resulting anisotropy would be small for
Cu Benzoate and is neglected in the present letter.

When an external magnetic field is present, the Zeeman term appears
in the original Hamiltonian. For example, if we apply the magnetic
field in $x$-direction, the Zeeman term is
$H_{\rm Zeeman} = - H_0 \sum_j S^x_j $ where $H_0$ is the external field
After the redefinition~(\ref{eq:redef}), the Zeeman term gives
\begin{equation}
  - H_0 \cos{\frac{\alpha}{2}} \sum_j S^x_j
  - H_0 \sin{\frac{\alpha}{2}} \sum_j (-1)^j S^y_j .
\label{eq:rZeeman}
\end{equation}
Namely, the effective staggered field of strength $H_0 \sin{(\alpha/2)}$
is generated in $y$ direction. 
For general directions of the DM vector $\vec{D}$
and the external field $\vec{H}_0$,
the direction of the effective staggered field is $\vec{H}_0 \times \vec{D}$.
If $D << J$, then $\alpha \sim D/J$ and
the staggered field is given by $ \vec{H}_0 \times \vec{D} /2$.

In Cu Benzoate, a staggered field is already present
before the redefinition, due to the alternating $g$-tensor.
The total effective staggered field is obtained by the redefinition
of the Zeeman term~(\ref{eq:rZeeman})
together with the alternating $g$-tensor. For a small
alternating part of the $g$-tensor, the total effective staggered
field is given by a sum of two effects.
Both effects produces a transverse staggered field
(orthogonal to the direction of the applied field),
apart from the small longitudinal component due to the
uniform part of the off-diagonal elements of the $g$-tensor.
We neglect the longitudinal component of the staggered field,
which is actually very small in Cu Benzoate.

Thus we are led to consider a one dimensional Heisenberg Hamiltonian
with mutually perpendicular uniform field $H$ and staggered field $H$
\begin{equation}
\hat{H} =\sum_i[J\vec S_i\cdot \vec S_{i+1} -HS^x_i+h(-1)^iS^z_i],
\label{eq:modelham}
\end{equation}
with $H >> h$.
It is instructive to
analyse the model~(\ref{eq:modelham})
in the standard spin-wave theory approximation
(lowest order $1/s$ expansion)~\cite{SakaiShiba},
although some of the conclusions
will be modified when we take into account one-dimensional quantum
fluctuations more accurately.
The classical groundstate is a
canted antiferromagnetic structure.  The canting
angle measured from the $x$-axis, $\theta$,
determined by minimizing the classical energy is
the solution of:
$2Js\sin 2\theta -H\sin \theta -h\cos \theta =0$.
Now considering fluctuations around this classical groundstate to
lowest order in $1/s$ gives 2 branches of spin-waves, in the
antiferromagnetic Brillouin zone ($|k|<\pi /2$) with energies:
\begin{eqnarray}
E_{\pm}(k) &=& 
  \{[2Js\cos{2\theta} +H \sin{\theta} + h \cos{\theta}
    \pm Js(1- \cos{2\theta})
\nonumber \\
&&
\cos{k}]^2 -[Js(\cos{2\theta} +1)\cos{k}]^2\}^{1/2}.
\end{eqnarray}
In the case $h=0$, the minimum energy of two modes are
$E_+=H$ and $E_-=0$ at $k=0$.  The $-$ Goldstone mode corresponds
to a precession of the spins around the $x$ axis.  A non-zero
staggered field $h$ gives this mode a finite gap.  To leading order in
$h$ but all orders in $H$ this is given by:
\begin{equation}
E_-(0)=\sqrt{4Jsh [1+(H^2/8J^2s^2)]} [1-(H/4Js)^2]^{1/4}.
\end{equation}
Note the singular dependence on $h$ but the weak dependence on $H$; $E_-(0)$
is essentially independent of $H$ until $H\approx O(Js)$. 
Conversely, the upper mode $E_+(0)$ depends only weakly on $h$ but
strongly on $H$. 
In the case $h=0$, the existence of the upper mode at energy $H$ is
more rigorously established~\cite{Mueller:field}
without the spin-wave approximation.
The upper mode $E_+$ is presumably observed in the
experiment: the peak at higher energy $\hbar \omega \sim 0.8 {\rm meV}$
in Fig. 2(a) of Ref.~\cite{Dender:s12} is consistent with
the upper mode energy $H = 7 {\rm T} \sim 0.8 {\rm meV}$.
Taking into account 1D critical fluctuations, the power-law
behaviour of the lower gap $E_-(0)\propto h^{1/2}$ is changed to $h^{2/3}$,
as we will discuss below.
It is reasonable to expect the weak dependence of the lower gap on $H$ to
remain true.  

The low-energy behavior of the system should be well described by
Abelian bosonization.
In the bosonization approach, the only effects of the uniform field $H$
is shift of the Fermi momentum $k_F$ and the renormalization of the
compactification radius $R$.
The transverse staggered field is mapped to
the operator $\cos{( 2 \pi  R \tilde{\phi})}$.
Thus, the effective low-energy theory for the model~(\ref{eq:modelham})
is given by the sine-Gordon model with the Lagrangian density
\begin{equation}
  {\cal L} = \frac{1}{2} ( \partial_{\mu} \phi )^2 
           + \lambda \cos{(2 \pi R \tilde{\phi})}.
\label{eq:sGlag}
\end{equation}
The operator $\cos{(2 \pi R \tilde{\phi})}$
has dimension $\pi R^2$, and is more relevant than
the exchange anisotropy. Actually, it is the most relevant operator
in the system.
At zero field, the system is isotropic and $R=1/\sqrt{2 \pi}$);
the dimension of the staggered field operator is $1/2$.
The radius $R$ is affected mainly by the uniform field.
Its effect may be estimated from the Bethe Ansatz solution for
the Heisenberg model under a uniform field.
The dimension of the operator $\cos{(2 \pi R \tilde{\phi})}$
is reduced to $\pi R^2 = 0.41$ at $H \sim 0.52J$ ($7$ T for Cu Benzoate).
While the uniform field does affect the low-energy excitation, 
the effect is not drastic. This is consistent with the spin-wave calculation.

A Renormalization-Group argument shows that the gap $\Delta$ is proportional
to $h^{1/(2-d)}$ where $d$ is the dimension of the relevant operator
and $h$ is the total effective staggered field.
If we neglect the effect of the uniform field $H$,
$d=1/2$ and thus $\Delta \sim h^{2/3} = h^{0.67}$.
Precisely speaking, there is a log-correction due to the presence of
the marginal operator~\cite{AGSZ}:
$\Delta \sim h^{2/3} |\log{h}|^{1/6}$.
The log-correction is not significant in the present case.
While the field-theory argument gives the exponent for the gap,
it does not determine the magnitude.
Thus, we  studied numerically the excitation gap of the $S=1/2$
Heisenberg antiferromagnetic chain under a staggered field (but
with no uniform field), by Lanczos method up to 22 sites.
The result is shown in Fig.~1. 
We found that, for small staggered field $h$, the lowest excitation
gap to the total magnetization $1$ sector behaves consistently
with the field-theory prediction.
We fixed the proportionality constant as
\begin{equation}
  \Delta = 1.85 (\frac{h}{J})^{2/3} J | \log{\frac{h}{J}} |^{1/6} 
\label{eq:gapcoef}
\end{equation}
from the numerical result.

\begin{figure}
\begin{center}
\leavevmode
\epsfxsize=3.5in
\epsfbox{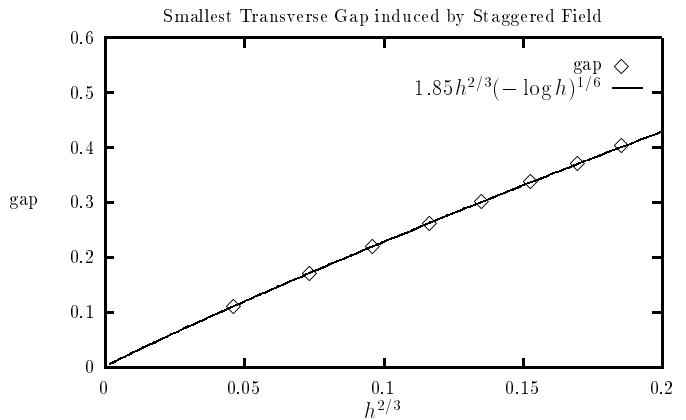}
\end{center}
\caption{The lowest excitation gap to $\sum S^z =1$ sector
in the Heisenberg antiferromagnetic chain,
induced by the staggered field $h$.
The gap is obtained by an extrapolation of finite-size gap
by Lanczos method upto 22 sites.
Both gap and $h$ are measured in unit of the coupling constant $J$.
The data is well fit by the field-theory prediction
$h^{2/3} | \log{h}|^{1/6}$, with a coefficient $1.85$.}
\end{figure}

Since the effective staggered field $h$ is proportional to the applied
field $H_0$, the gap should be proportional to ${H_0}^{2/3}$.
This is in a good agreement with the experiment~\cite{Dender:s12}
in which the gap is found to scale as ${H_0}^{0.65(3)}$ for three
directions of the magnetic field. This supports our basic claim
that the field-induced gap is due to the effective staggered field.
If we include the change in $R^2$ due to the uniform field, the
exponent changes to $0.63$ at $H_0 =7 {\rm T}$.
Taking an average, the agreement with the experiment is improved.
Moreover, the sine-Gordon model~(\ref{eq:sGlag}) is integrable.
The elementary excitations are given by soliton, anti-soliton and
soliton-antisolition boundstates (``breathers''), and their exact
mass ratios are available~\cite{Dashen:sGmass,Zam2:Smatrix}
(For an introduction, see for example Ref.~\cite{Tsvelik:QFT}.)
At the isotropic point $R=1/\sqrt{2 \pi}$, there are two kinds of
breathers, and the mass of the lighter breather is degenerate with
the soliton/anti-soliton, forming a triplet.
The mass ratio of the triplet and the singlet (heavier breather)
is $1: \sqrt{3}$.
When the $SU(2)$ symmetry is broken, the triplet is split and
the mass ratio for light breather, (anti-)soliton and heavy breather is
$2 \sin{[ \pi^2 R^2 / (4 - 2 \pi R^2)]} : 1 :
2 \sin{[ \pi^2 R^2 / (2 - \pi R^2)]}$. 
For $H_0 = 7 T$, $\pi R^2 = 0.41$ and the ratio is $0.79 : 1 : 1.45$.
This is close to the ratio of three peaks observed in Figure 2
of Ref.~\cite{Dender:s12}, at $0.17$ meV, $0.22$ meV and
$0.34$ meV. ($0.77:1:1.55$).

In the experiment, the magnitude of the
staggered field due to the alternating $g$-tensor
depends on the direction of the applied field.
The ratio is $0.019:0.053:0.049$ for field applied
in $a'',b,c''$ directions. 
This can be understood in our theory,
because the proportionality constant between
$h$ and $H_0$ depends on the field direction.
If we only consider the staggered $g$-tensor effect, the ratio of the gap is
$0.019^{2/3}: 0.053^{2/3} : 0.049^{2/3} \sim 1:2.0:1.9 $.
This does not explain the observed ratio of the
gap $0.55:1:2.0$. In particular, the order of
gap for $b$ and $c''$ is reversed.
Thus, it is necessary to include the effective staggered field due to
the DM interaction, in order to explain the gap.

In general, the magnitude of $\vec{D}$ is argued to be of order of
$(\Delta g /g) J$ where $\Delta g$ is a shift of $g$-factor in the
crystal~\cite{Moriya:DMint}.
In Cu Benzoate, $\Delta g / g \sim 0.1$.
While more precise estimate of $\vec{D}$ in Cu Benzoate is unknown,
it should be in $ac$-plane (or equivalently $a''c''$-plane) from the
crystal structure~\cite{Moriya:DMint,Oshima:CuBenzAFR}.
Thus $\vec{D}$ is specified by two parameters, for example,
by $D = | \vec{D} |$ and the angle $\chi$ between $\vec{D}$ and $a''$-axis.
We first determined $\vec{D}$ so that it reproduces the experimentally
observed angle dependence of the gap $a'':b:c'' = 0.55:1:2.0$.
We found two solutions: $( \chi, D) = (0.22,0.034J)$ and $(-0.0066, 0.10J)$.
($\chi$ is in radians.)
Both directions are close to $a''$-axis (or $a'$-axis) as claimed
in Ref.~\cite{Oshima:CuBenzAFR}.
Moreover, both values of $D$ are consistent with the general estimate
$D \sim (\Delta g /g) J \sim 0.1 J$. Thus, it has been shown that
a reasonable magnitude of the DM interaction
can give the angle dependence observed in the experiment.

We can also estimate the magnitude of the gap from our theory,
using~(\ref{eq:gapcoef}). 
For the former solution $( \chi, D) = (0.22,0.034J)$, the gap
for $H_0 =7 {\rm T}$ applied in $b$-direction is $0.096J$.
For the latter solution $(\chi, D) = (-0.0066,0.10J)$, the gap
for the same field is $0.15J$.
Both gives a correct order of magnitude compared to the experimental
value $\sim 0.2 {\rm meV} = 0.13J$ in Ref.~\cite{Dender:s12}.
More quantitative comparison would require further analysis of the specific
heat data, since aspects of the treatment in Ref.~\cite{Dender:s12} could be
questioned. They fit the low-temperature specific heat by
six independent massive bosons with same gap $\Delta$, but
the sine-Gordon theory predicts four elementary excitations with
different masses, as discussed.
While their estimate presumably gives correct order of magnitude,
the precise value would be changed by a refined analysis.
Experiments with other field directions would provide a further check
of our theory.

Finally, we comment on other consequences of our theory.
From a scaling argument, the staggered magnetization
behaves as $H_0^{1/3}$.
The direction of the staggered magnetization is given by
the effective staggered field. Thus our theory could be tested if
the staggered magnetization is measured.
Moreover, by the redefinition~(\ref{eq:redef}),
the physical spin operator corresponds to a
rotated spin operator in the Heisenberg antiferromagnet
without the DM interaction.
While it has no drastic effect on the neutron scattering
experiment, it affects the susceptibility measurement dramatically.
The observed susceptibility $\chi_{\rm exp}$ is
given by a linear combination of the uniform susceptibility
$\chi_u$ and the staggered one $\chi_s$ of the Heisenberg model. 
Since the latter diverges at low temperature,
$\chi_{\rm exp}$ would also diverge at low temperature.
This could explain the enhancement of the susceptibility
observed in Ref.~\cite{Dender:CuBenzchi}, though a quantitative
theory would require inclusion of interchain interactions.
Further discussions, including details of the arguments in the
present letter, will be given in a future publication.

We thank Collin Broholm for many stimulating discussions,
as well as for providing results prior to publication.
The numerical calculation in this work was based on
the program package TITPACK ver 2.0 by H. Nishimori.

\end{document}